\documentclass{ws-procs9x6}

\usepackage{graphicx}
\usepackage{dcolumn}
\usepackage{bm}

\begin{document}

\title{Contribution of Instanton Induced Interaction for Penta-quarks \\in MIT Bag Model}

\author{Tetsuya Shinozaki and Makoto Oka}
\address{%
Department of Physics, H-27, Tokyo Institute of Technology,\\ Meguro, Tokyo 152-8551, Japan\\
E-mail: shinozk@th.phys.titech.ac.jp
}%
\author{Sachiko Takeuchi}%
\address{%
Japan College of Social Work, Kiyose 204-8555, Japan
}%

\maketitle

\abstracts{
Roles of instanton induced interactions (III) in the masses of pentaquark baryons,  $\Theta^+$ ($J=1/2$ and 3/2) and $\Xi^{--}$, and a dibaryon, $H$, are discussed using the MIT bag model. 
It is shown that the two-body terms in III give a strong attraction mainly due to the increase of the number of pairs in multi-quark systems. In contrast, the three-body $u$-$d$-$s$ interaction is repulsive.
It is found that III lowers the mass of negative-parity $\Theta^+$ as much as 100 MeV from the mass predicted by the bag model without III.
}

Reports of discoveries\cite{Nakano:2003qx} of exotic baryons started intensive discussions on various possibilities of bound pentaquark states. The most essential properties are spin and parity.  Predictions  by the chiral soliton model\cite{Diakonov:1997mm} and various quark models\cite{Karliner:dt,Jaffe:2003sg,Kochelev:2004nd} claimed a $1/2^{+}$ state of $\Theta^+$, while constituent quark model with all the five quarks sitting in the lowest energy level predicts a negative parity ground state, $1/2^{-}$.
Furthermore, majority of QCD-based calculations, such as QCD sum rule\cite{Sugiyama:2003zk} and lattice QCD\cite{Ishii:2004qe}, indicate that the positive parity state has a higher mass.

The constituent quark model has several important dynamical ingredients, among which we here consider confinement, perturbative one-gluon exchange (OgE) interaction and nonperturbative instanton induced interaction (III). Interesting roles of III in the baryon spectrum were studied\cite{Shuryak:bf,Oka:ud}.

The purpose of this paper is to clarify roles of III in the pentaquark systems in the context of the MIT bag model.
In the naive bag model, the ``vacuum'' inside the bag is identified as the perturbative vacuum.
Therefore we consider a situation that the vacuum structure is minimally modified to reproduce appropriate spectrum of the pseudoscalar mesons. 
On the other hand, it is expected that the confinement mechanism that successfully describes three-quark baryon states is common to the pentaquark baryons. Also there are some advantages of the bag model. The quarks inside the baryon are treated relativistically. The size of the baryons can be automatically determined from pressure balance at the bag surface.

The instanton induced interaction (III), introduced by 't Hooft\cite{'tHooft:fv}, is an interaction among quarks of $N_f\,(=3)$ light flavors,
The main difference from the perturbative gluon-exchange interactions is that III is not chirally invariant and applies only on flavor singlet states of quarks. The interaction is written as a contact interaction\cite{Oka:1990vx}.
\begin{eqnarray}
H^{(3)}&=&G^{(3)} \epsilon_{ijk} \epsilon_{i'j'k'} \, \bar{\psi}_{R,i}(1)\bar{\psi}_{R,j}(2)\bar{\psi}_{R,k}(3) \left( 1-\frac{1}{7} \sum_{i<j} \sigma_i \cdot \sigma_j \right)\nonumber\\ &\times& \psi_{L,k'}(3)\psi_{L,j'}(2)\psi_{L,i}(1) + \mathrm{(h.c.)}, \nonumber\\
H^{(2)}&=&G^{(2)} \epsilon_{ij} \epsilon_{i'j'}  \, \bar{\psi}_{R,i}(1)\bar{\psi}_{R,j}(2) (1-\frac{1}{5}\sigma_1 \cdot \sigma_2)\psi_{L,j'}(2)\psi_{L,i'}(1) + \mathrm{(h.c.)}, \nonumber
\end{eqnarray}
where $\epsilon_{ijk}$ is totally asymmetric tensor, and $i$, $j$ and $k$ represent flavor.  $R$ and $L$ are chiral indices. $H^{(3)}$ is the 3-body Hamiltonian and $H^{(2)}$ is  the 2-body Hamiltonian obtained by contracting a quark pair in the 3-body III into a quark condensate:
$G^{(2)}_{ud} = \frac{25}{7} \frac{\langle \bar{u} u \rangle}{2}\frac{m_s^{\mathrm{eff}}}{m_u^{\mathrm{eff}}} G^{(3)}$,
$G^{(2)}_{us} = m_u^{\mathrm{eff}}/m_s^{\mathrm{eff}} G^{(2)}_{ud}$.
where $m^{\mathrm{eff}}$ is the constituent mass of the quark. We use $m_u^{\mathrm{eff}}/m_s^{\mathrm{eff}} \simeq 0.6$. The $\langle \bar{u} u \rangle \simeq (-225\mathrm{MeV})^3$ is the quark condensate.
The 3-body interaction $H^{(3)}$ is  repulsive,  while the 2-body interaction $H^{(2)}$ is attractive  because the quark condensate is negative. 
Some previous works consider III in the context of diquark models of the pentaquark baryons\cite{Kochelev:2004nd}.

The mass of a hadron in the MIT Bag Model\cite{DeGrand:cf} is given by 
\begin{eqnarray}
M(R) &=& n_u w(m_u,R) + n_s w(m_s,R)+ 4\pi/3 \, B R^3 - Z_0/R \nonumber\\&+& (1 - P_{III}) \sum_{i>j} (\vec{\sigma_i} \cdot \vec{\sigma_j})\,(\vec{\lambda_i}\cdot \vec{\lambda_j}) M_{ij}(R) \nonumber\\ &+& P_{III}(H^{(3)}(R)+H^{(2)}(R)) + E_0 ,\nonumber
\end{eqnarray}
where $R$ is the bag radius, The fifth term is the color-magnetic part from OgE. 
$P_{III}$ is a parameter which represents the portion of the hyperfine splitting induced by III. 
If $P_{III}=0$, the mass splitting of $N-\Delta$ comes purely from OgE, while for $P_{III}=1$ it comes purely from III.
The way of determining $P_{III}$ is to reproduce the $\eta-\eta'$ mass difference. In the case of the MIT Bag model\cite{Takeuchi:sg},
We estimated around $P_{III}=0.3$.
$E_0$ is introduced to reproduce the mass of the nucleon.
It is given roughly by $E_0 = 150 \mathrm{MeV} \times P_{III}$.
The $E_0$ can be taken into account by changing $Z_0$ and $B$ accordingly, but here we remain to fix $Z_0$ and $B$.
The other parameters of the bag model are taken from the original MIT bag model\cite{DeGrand:cf}.

We consider the pentaquarks  $\Theta^{+}$ composed of  $uudd\bar{s}$ with isospin 0, spin 1/2 and negative parity, 
and $\Xi^{--}$, a partner within the flavor $\bar{10}$ with isospin 3/2. 
We also consider $\Theta^{+}_{S=3/2}$, which is the spin partner of $\Theta^{+}$. 

\begin{figure}
\begin{center}
\includegraphics[width=0.495\textwidth]{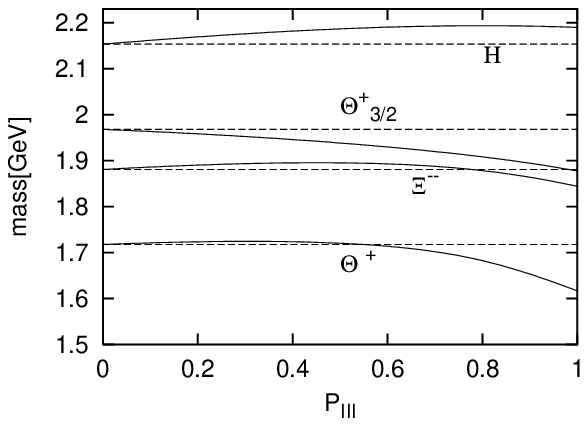}
\includegraphics[width=0.495\textwidth]{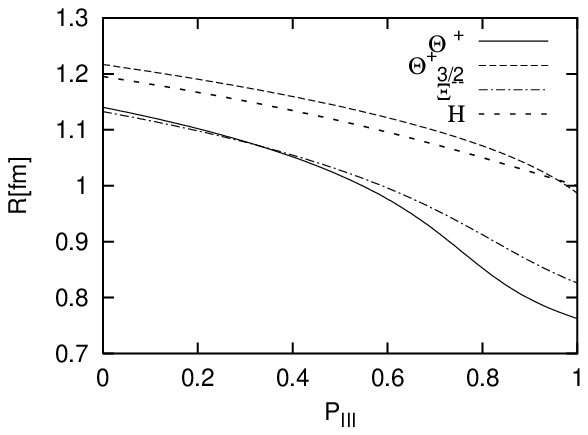}
\end{center}
\caption{\label{mass}The masses (left) and  the bag radii (right). }
\end{figure}

In Fig. \ref{mass},  we show the masses of the pentaquarks as functions of $P_{III}$.
The dashed lines are the values at $P_{III}=0$, which correspond to the masses under the influence only of OgE.
The right end, $P_{III}=1$, gives the masses when the $N-\Delta$ splitting is purely due to III. 
We point out that the pure OgE lowers the masses of $\Theta^{+}$ from the noninteracting 5 quark state.
One sees that the $\Theta^{+}$ is affected by III most strongly among these states.
At close to the $P_{III}$=1, the mass of $\Theta^{+}$ is 100MeV smaller than that at $P_{III}=0$.
In contrast, $\Theta^{+}_{S=3/2}$ changes significantly in all $P_{III}$.
But the mass of $\Xi^{--}$ is almost constant. The mass of the $H$ dibaryon grows monotonically as $P_{III}$ increases.
It is found that the mass of $\Theta^{+}$ does not agree with the experimental value (1540MeV) even if the full III is introduced.
On the other hand, the model reproduces the mass of $\Xi^{--}$. 
The contribution of the 3 body III is roughly 10\% of that of the 2 body III for the pentaquarks.
For the $\Theta^{+}_{S=3/2}$, the contribution of OgE is very small.
Thus effects of III are most easily seen in $\Theta^{+}_{S=3/2}$. 

In Fig. \ref{mass}, the radii of the considered baryons are given.
They show that $\Theta^{+}$  shrinks as $P_{III}$ increases. 
At the realistic region , $P_{III}=0.3$, the radii of pentaquarks are about $0\%\sim20\%$ larger than the radius of the nucleon, 5 $\mathrm{GeV}^{-1}$. We find that the strongly attractive force of III makes the bag radii shrink. In fact, the radii of the pentaquarks are as small as the radius of the 3-quark baryons. 

We conclude that the effects of III  have been studied using the MIT bag model in the negative parity case.
We have found that III lowers the mass of $\Theta^+$ and $\Theta^{+}_{S=3/2}$, while the mass of $H$ increases as the strength of III increases.  The present results can not reproduce the observed $\Theta^+$ mass. Possible resolutions are corrections from expected two-body (diquark type) correlations, pionic effects, which may be included in chiral bag models, and also couplings to background $NK$ scattering states.  If these effects are important, the pentaquark spectrum may be well modified. Despite these defects, the current study is worthwhile because using the simplest possible picture of the hadron, we demonstrate how large and important are the effects of instantons on the spectrum of pentaquarks. Further analysis including the above-mentioned corrections are to be performed as the next step.

\section*{Acknowledgments}
This work is supported in part by the Grant for Scientific Research (B)No.15340072, (C)No.16540236 and (C)No.15540289 from the Ministry of Education, Culture, Sports, Science and Technology, Japan. T.~S. is supported by a 21st Century COE Program at Tokyo Tech "Nanometer-Scale Quantum Physics" by the Ministry of Education, Culture, Sports, Science and Technology.

\end{document}